\documentclass[epj]{svjour}

\usepackage{amsmath,amssymb,amsfonts}
\usepackage{graphicx,subfigure}
\usepackage{cite}
\usepackage{graphics}

\newcommand{\sign}{\text{sign}} \newcommand{\erf}{\text{erf}}
 
 \newcommand{\prob}{\text{Prob}}
\newcommand{\avg}[1]{\left\langle{#1}\right\rangle}

\newcommand{\disavg}[1]{\left[{#1}\right]_{\text{dis}}}
\newcommand{\pathavg}[1]{\left\langle{#1}\right\rangle_{\text{paths}}}
\newcommand{\timeavg}[1]{\left\langle{#1}\right\rangle_{\text{time}}}
\newcommand{\zavg}[1]{\left[{#1}\right]_{\boldsymbol{\zeta}}}

\newcommand{\ovl}[1]{\overline{#1}} 
\newcommand{\ii}{\text{i}}

\renewcommand{\l}{\left}
\renewcommand{\r}{\right}
\begin{document}

\title{Dynamics of multi-frequency minority games}

\author{Andrea De Martino}

\institute{INFM--SMC and Dipartimento di Fisica, Universit\`a di Roma
``La Sapienza'', p.le A. Moro 2, 00185 Roma
(Italy)\\\email{andrea.demartino@roma1.infn.it}}

\date{~}

\abstract{The dynamics of minority games with agents trading on
different time scales is studied via dynamical mean-field theory. We
analyze the case where the agents' decision-making process is
deterministic and its stochastic generalization with finite
heterogeneous learning rates. In each case, we characterize the
macroscopic properties of the steady states resulting from different
frequency and learning rate distributions and calculate the
corresponding phase diagrams. Finally, the different roles played by
regular and occasional traders, as well as their impact on the
system's global efficiency, are discussed.
} 
%

\titlerunning{Dynamics of multi-frequency minority games}

\maketitle
\section{Introduction}

The collective phenomena that characterize the evolution of
competitive populations of adaptive agents, such as the onset of
cooperation or the creation of exploitable information, have attracted
a great deal of attention from statistical physicists over the past
few years. The hope is that the occurrence of such macroscopic effects
can be understood starting from the laws that govern the behavior of
the individual agents. Thanks to the basic simplicity of its
definition, the minority game (MG) \cite{cz1,web} allows a full
theoretical analysis of many of these issues. Originally, it was
designed to mimic a market of speculators subject to the law of supply
and demand. At each time step, traders react to the receipt of a
public information pattern (the `state of the world') by either buying
or selling, trying to profit from price fluctuations to maximize their
payoff, and learning from experience. Upon decreasing the relative
number $\alpha$ of possible information patterns, the degree of
cooperation in the system, measured by the inverse magnitude of global
fluctuations, increases, until a critical point $\alpha_c$ is reached,
below which highly cooperative as well as highly uncooperative states
can occur, depending on the initial conditions of the agents' learning
processes. Remarkably, the low $\alpha$ phase is characterized by the
absence of exploitable information, while in the high $\alpha$ phase
the market is to some degree predictable and hence inefficient.

On the technical level, the MG is defined by a set of globally coupled
zero-temperature Markov processes with quenched disorder and without
detailed balance. The system at $\alpha_c$ undergoes a dynamical phase
transition with ergodicity breaking, related to the onset of anomalous
response. The stationary state is solvable, at least in the ergodic
phase, via both static (replica) and dynamical methods, so that a very
deep understanding of its macroscopic properties is
achievable. Moreover, the MG is open to a large class of economically-
and biologically-inspired extensions, that allow to tackle such key
issues as the interplay of different types of traders in a market
\cite{web}.

An interesting modification of the original model is obtained when
synchronicity is removed, and one accounts for the possibility that
agents trade on different time scales. This is the `colored' minority
game (CMG) first introduced in \cite{mp}. Assuming a power-law
distribution of trading frequencies and using the static approach of
\cite{prl,mcz}, it was shown that, while the phase transition picture
is substantially preserved, the actual critical point $\alpha_c$
depends on the particular exponent entering the power
law. Furthermore, regular and occasional traders were shown to take
markedly different behaviors, with the latter less prone to change
trading strategy and hence contributing significantly less to the
growth of global fluctuations.

In this work, we adapt the generating functional methods originally
devised for spin glasses \cite{dedo,kt,crhs,ck,bckm} and already
employed in the theory of MGs \cite{hc,hd,chs,gcs,dgm} to study the
dynamics of the CMG in a few different cases. First, assuming that the
agents' learning process is deterministic, we derive a stochastic
equation describing the behavior of an effective agent trading with
frequency $f$ (Sec.~2). Global dynamical quantities such as
correlation and response functions appear as proper averages over the
trading frequency distribution $q(f)$. We calculate the phase diagrams
exactly and evaluate macroscopic order parameters in ergodic
stationary states explicitly in the simple case where $q(f)$ is
bimodal (i.e. for a mixed population of frequent and occasional
traders, with a wide separation of time scales between the former and
the latter, Sec.~3) and when it has a power-law form (i.e. when there
is no single characteristic time scale for trading, Sec.~4). The roles
of regulars and irregulars can be thoroughly investigated. Then
(Sec.~5), we move to the case where the agents' learning process is
stochastic by introducing different learning rates as a further
element of heterogeneity among agents (besides their strategies and
trading frequencies). We analyze additive and multiplicative decision
noise (Sec.~6), deriving the corresponding phase structure as a
function of the learning rates. Finally, we summarize our results and
formulate our conclusions (Sec.~7). In order to keep the treatment
concise, we will not report the full generating functional analysis
and limit ourselves to a study of the resulting effective mean-field
equations. The reader is from this moment referred to \cite{hc,chs}
for all details about its derivation in similar instances.

\section{CMG with deterministic decision-making}

The basic ingredients of the MG are as follows. There are
$N$ agents. At each round $n$ of the game all agents receive the same
information pattern $\mu(n)$ drawn at random with uniform probability
from a set of $P$ possible. $P$ is assumed to scale linearly with $N$,
so that in the limit $N\to\infty$ the parameter $\alpha=P/N$ (the
relative number of information patterns) remains finite. We assume
that each agent is endowed with two different strategies (labeled by
$g=1,2$) to convert the information into a ternary trading decision:
$\boldsymbol{a}_{ig}:\{1,\ldots,P\}\ni\mu\to a_{ig}^\mu\in\{-1,0,1\}$.
$a_{ig}^\mu$ is the action prescribed to agent $i$ by his $g$-th
strategy upon receipt of information $\mu$. One might think that
$a_{ig}^\mu=\pm 1$ correspond to `buy' and `sell', respectively, while
$a_{ig}^\mu=0$ stands for `do nothing'. For all $i$, $g$ and $\mu$,
each $a_{ig}^\mu$ is selected randomly and independently from
$\{-1,0,1\}$ before the start of the game and fixed. Following
\cite{mp}, we take a probability distribution of the form
\begin{equation}\label{pd}
P(a_{ig}^\mu)=\frac{f_i}{2}\delta_{a_{ig}^\mu,1}+
\frac{f_i}{2}\delta_{a_{ig}^\mu,-1}+(1-f_i)\delta_{a_{ig}^\mu,0}
\end{equation}
The numbers $f_i$ ($0\leq f_i\leq 1$) represent the trading
frequencies of agents. They can be seen as an additional family of
i.i.d. quenched random variables with a prescribed probability density
which we denote by $q(f)$. In the original MG, where all
agents trade at all rounds, one has $q(f)=\delta(f-1)$.

The system's dynamics is defined through the microscopic stochastic
equations that govern the decision making of the individual
traders. Each strategy of every agent is given an initial valuation
$p_{ig}(0)$, which is updated at the end of every round. Loosely
speaking, $p_{ig}(n)$ measures the performance of $g$ up to round
$n$. At the beginning of round $n$, every trader selects his so-far
best-performing strategy, $\widetilde{g}_i(n)=\arg\max p_{ig}(n)$, and
formulates a bid according to the trading decision it prescribes:
$b_i(n)=a_{i\widetilde{g}_i(n)}^{\mu(n)}$. The total bid (or, the
excess demand) at round $n$ is simply $A(n)=(1/\sqrt{N})\sum_{i=1}^N
b_i(n)$. Once $A(n)$ is known, strategy valuations are updated
according to
\begin{equation} \label{perfdyna}
p_{ig}(n+1)=p_{ig}(n)-a_{ig}^{\mu(n)}A(n)
\end{equation}
and agents move to the next round. Strategies that would have
prescribed the minority (resp. majority) action, i.e. such that
$a_{ig}^{\mu(n)}A(n)<0$ (resp. $>0$), are thus rewarded
(resp. penalized). The valuation of strategies with
$a_{ig}^{\mu(n)}=0$ is instead left unchanged. We assume that agents
neglect their market impact \cite{mcz}.

Introducing the `preferences' $y_i(n)=[p_{i1}(n)-p_{i2}(n)]/2$, and
the quantities
$\boldsymbol{\xi}_i=(\boldsymbol{a}_{i1}-\boldsymbol{a}_{i2})/2$,
$\boldsymbol{\omega}_i=(\boldsymbol{a}_{i1}+\boldsymbol{a}_{i2})/2$,
and $\boldsymbol{\Omega}=(1/\sqrt{N})
\sum_{i=1}^N\boldsymbol{\omega}_i$, (\ref{perfdyna}) becomes
\begin{equation}\label{dyna}
y_i(n+1)= y_i(n)-\xi_i^{\mu(n)}[\Omega^{\mu(n)}+\frac{1}{\sqrt{N}}\!
  \sum_{j=1}^N\xi_j^{\mu(n)}s_j(n)]
\end{equation}
The Ising spin $s_i(n)=\sign[y_i(n)]\in\{-1,1\}$ is the relevant
dynamical variable: $s_i(n)=1$ (resp. $-1$) indicates that agent $i$
has chosen strategy $g=1$ (resp. $2$) at round $n$.

The simplest dynamical approach to the analysis of the stationary
states of (\ref{dyna}) is based on the `batch' approximation, first
employed in \cite{hc}. It consists in averaging (\ref{dyna}) over
$\mu$. The result is
\begin{equation}\label{batch}
y_i(t+1)=y_i(t)-h_i-\sum_{j=1}^N J_{ij}s_j(t)
\end{equation}
where $t$ is a rescaled time,
$h_i=(2/\sqrt{N})~\boldsymbol{\Omega}\cdot\boldsymbol{\xi}_i$ and
$J_{ij}=(2/N)~\boldsymbol{\xi}_i\cdot\boldsymbol{\xi}_j$. This choice
describes strictly speaking a situation in which valuation updates are
made once every $P$ steps or, equivalently, considering the average
effect of all possible $\mu$'s. It is now well known, that the
stationary state of (\ref{batch}) is qualitatively and quantitatively
very similar, though not identical, to that of (\ref{dyna}). The
analysis of (\ref{batch}) is however much simpler and more
straightforward than that of the original `on-line' dynamics
(\ref{dyna}).

In the limit $N\to\infty$, the Markovian multi-agent process
(\ref{batch}) can be studied \`a la De Dominicis \cite{dedo} by
introducing the generating functional of the dynamics, i.e.
\begin{multline}
Z[\boldsymbol{\psi}]=\pathavg{e^{\ii\sum_{it}y_i(t)\psi_i(t)}}=\\
=\int p(\boldsymbol{y}(0))~ e^{\ii\sum_{it}\widehat{y}_i(t)
[y_i(t+1)-y_i(t)-\theta_i(t)]+y_i(t)\psi_i(t)}\times\\ \times
e^{\ii\sum_{it}\widehat{y}_i(t)[h_i+\sum_j J_{ij}s_j(t)]}
\prod_{it}[dy_i(t)d\widehat{y}_i(t)/(2\pi)]
\end{multline}
and carrying out the average over the quenched disorder (i.e. over the
$\boldsymbol{a}_{ig}$'s with probability distribution
(\ref{pd})). This procedure ultimately results in a non-Markovian
stochastic equation for a single effective agent whose properties are
equivalent to those of the original $N$-agent system. In our case, the
effective agent is trading with frequency $f$. As the calculation is
standard, we will limit ourselves to the final outcome, referring the
reader to e.g. \cite{hc} for details of a closely related case. The
effective-agent equation turns out to be given by
\begin{equation}\label{esap}
y(t+1)=y(t)+\sum_{t'\leq t}R(t,t')s(t')+\theta(t)+\sqrt{\alpha f}~z(t)
\end{equation}
where $R(t,t')$ describes the retarded self-interaction, whose precise
form will be given below, $\theta(t)$ is an external field added
to probe the system against small perturbations, and $z(t)$ is a
zero-average Gaussian noise with correlations
$\avg{z(t)z(t')}=\Lambda(t,t')$, with
\begin{equation}\label{corr}
\mathsf{\Lambda}=[(\mathsf{I}+\mathsf{G})^{-1}(\mathsf{F}+\mathsf{C})
(\mathsf{I}+\mathsf{G}^T)^{-1}]
\end{equation}
We used the following notation: $\mathsf{I}$ stands for the identity
matrix, with elements $I(t,t')=\delta_{tt'}$; $\mathsf{F}$ denotes the
matrix with elements $F(t,t')=\ovl{f}$, with $\ovl{f}$ the average
trading frequency; $\mathsf{C}$ and $\mathsf{G}$ are defined by their
respective components
\begin{equation}
C(t,t')=\avg{s(t)s(t')}_*\quad\text{and}\quad
G(t,t')=\avg{\frac{\partial s(t)}{\partial\theta(t')}}_*
\end{equation}
where the $\avg{\cdots}_*$ average is intended over the effective
process (\ref{esap}) further averaged over $f$:
\begin{multline}\label{star}
\avg{\cdots}_*=\int df f q(f)\times\\
\times\int\prod_t\frac{dy(t)d\widehat{y}(t)}{2\pi}~e^{-\frac{\alpha
f}{2}\sum_{tt'}\widehat{y}(t)\Lambda(t,t')\widehat{y}(t')}\times\\
\times (\cdots)~e^{\ii\sum_t
\widehat{y}(t)[y(t+1)-y(t)-\theta(t)-\sum_{t'}R(t,t')s(t')]}
\end{multline}
$q(f)$ is the common probability density of the trading
frequencies. As usual, $C(t,t')$ and $G(t,t')$ can be identified with
the disorder-averaged correlation and response functions of the
multi-agent process:
\begin{gather}
C(t,t')=\frac{1}{N}\sum_i\disavg{\pathavg{s_i(t)s_i(t')}}\\
G(t,t')=\frac{1}{N}\sum_i\disavg{\frac{\partial}{\partial\theta_i(t')}
\pathavg{s_i(t)}}
\end{gather}
For simplicity of notation, we set
\begin{gather}
C(t,t')=\int f q(f) C(t,t'|f) df \label{opc}\\
G(t,t')=\int f q(f) G(t,t'|f) df \label{opg}
\end{gather}
Finally, the retarded self-interaction kernel has the form
\begin{equation}\label{rsi}
\mathsf{R}=-\alpha f (\mathsf{I}+\mathsf{G})^{-1}
\end{equation}
For $q(f)=\delta(f-1)$ the equations for the standard `batch' MG
\cite{hc} are immediately recovered.

Making for the asymptotic behavior of $\mathsf{C}$ and $\mathsf{G}$
the customary assumptions \cite{ckjpa} of time-translation invariance,
\begin{gather}\label{tti}
\lim_{t\to\infty}C(t+\tau,t)=C(\tau)\\
\lim_{t\to\infty}G(t+\tau,t)=G(\tau)
\end{gather}
finite integrated response,
\begin{equation}\label{fir}
\lim_{t\to\infty}\sum_{t'\leq t}G(t,t')<\infty
\end{equation}
and weak long-term memory
\begin{equation}\label{wltm}
\lim_{t\to\infty}G(t,t')=0~~~\forall t'\text{ finite}
\end{equation}
ergodic stationary states of the dynamics can be fully characterized
in terms of a few parameters. These are, in particular, the persistent
autocorrelation
\begin{equation}\label{cmax}
c=\lim_{\tau\to\infty}\frac{1}{\tau}\sum_{t<\tau}C(t)
=\int f q(f) c(f) df
\end{equation}
with $c(f)=\lim_{\tau\to\infty}(1/\tau)\sum_{t<\tau}C(t|f)$ the
autocorrelation of the effective trader (trading with frequency $f$),
and the susceptibility (or integrated response)
\begin{equation}\label{chimax}
\chi=\lim_{\tau\to\infty}\sum_{t\leq\tau}G(t)=\int f q(f) \chi(f) df
\end{equation}
where $\chi(f)=\lim_{\tau\to\infty}\sum_{t\leq\tau}G(t|f)$ is the
effective trader's susceptibility.

Following \cite{hc}, one can easily derive from (\ref{esap}) an
equation for the long-time behavior of the rescaled variable
$\widetilde{y}(t)=y(t)/t$:
\begin{equation}\label{longo}
\frac{\widetilde{y}}{\sqrt{\alpha f}}=-\gamma s+z,
~~~~~\gamma=\frac{\sqrt{\alpha f}}{1+\chi}
\end{equation}
where $\widetilde{y}=\lim_{t\to\infty}\widetilde{y}(t)$, while $s$ and
$z$ denote time-averages of $s(t)$ and $z(t)$, respectively. As in the
standard MG, part of the agents will `freeze' at $s=\pm 1$ and will
end up using just one of their strategies (for these,
$|y(t)|\to\infty$ asymptotically in such a way that $\widetilde{y}$
remains finite), whereas the remaining agents will keep flipping
between their strategies (for these, $y(t)$ remains asymptotically
finite and $\widetilde{y}$ is zero). It is easy to see that agents are
`frozen' for $|z|>\gamma$ while they are `fickle' (with $s=z/\gamma$)
for $|z|<\gamma$. The asymptotic values of the interesting quantities
can hence be immediately derived by separating the contributions of
frozen and fickle agents. For example, denoting by $\avg{~}$ an
average over the Gaussian r.v. $z$ with zero mean and variance (from
(\ref{corr})) $\avg{z^2}=(\ovl{f}+c)/(1+\chi)^2$ and defining
\begin{equation}
\phi(f)=\avg{\theta(|z|-\gamma)}=1-~\erf~\frac{\gamma}{\sqrt{2\avg{z^2}}}
\end{equation}
i.e. the fraction of agents trading with frequency $f$ that are frozen
(strictly speaking, the probability that the effective agent is
frozen), one has
\begin{multline}\label{cppp}
c(f)\equiv\avg{s^2}=
\avg{\theta(|z|-\gamma)}+\avg{(z/\gamma)^2\theta(\gamma-|z|)}=\\
=\phi(f)+\frac{\ovl{\phi}(f)}{\lambda(f)^2}-
\frac{1}{\lambda(f)}\sqrt{\frac{2}{\pi}}~e^{-\lambda(f)^2/2}
\end{multline}
with $\ovl{\phi}(f)=1-\phi(f)=\avg{\theta(\gamma-|z|)}$ and
\begin{equation}\label{lf}
\lambda(f)=\frac{\gamma}{\sqrt{\avg{z^2}}}=\sqrt{\frac{\alpha
f}{\ovl{f}+c}}
\end{equation}
Analogously, $\chi(f)$ can be calculated from the formula
\begin{equation}\label{chippp}
\gamma\sqrt{\alpha
f}~\chi(f)=\avg{\theta(\gamma-|z|)}=\ovl{\phi}(f)
\end{equation}
which is derived by noticing that the external field and the noise
term enter (\ref{esap}) in the same way, apart from the $\sqrt{\alpha
f}$ factor. The calculation of $\partial s/\partial z$ is then trivial
and leads to (\ref{chippp}), which in turn gives
\begin{equation}\label{chif}
\chi(f)=\frac{1+\chi}{\alpha f}~\erf~\frac{\lambda(f)}{\sqrt{2}}
\end{equation}

Inserting (\ref{cppp}) and (\ref{chif}) into (\ref{cmax}) and
(\ref{chimax}) one gets two equations for $c$ and $\chi$ which must be
solved self-consistently. Solutions will clearly depend on $\alpha$
\emph{and} on the underlying frequency distribution $q(f)$. It is the
purpose of the next two sections to study these equations and the
corresponding solutions in two interesting cases.

\section{Bimodal frequency distribution}

As a start, let us consider the simplest case, where
\begin{equation}
q(f)=(1-q)\delta(f-1)+q\delta(f-f_0)
\end{equation}
with $0\leq q\leq 1$ and $0\leq f_0\leq 1$, describing a mixed
population formed by frequent traders, who buy or sell at every round,
and occasional traders, who instead have a finite probability $1-f_0$
of taking no action. The population fractions of the two types are
$(1-q)$ and $q$, respectively.

The total fraction of frozen agents is evidently
\begin{multline}\label{phiphi}
\phi=(1-q)\phi(1)+q\phi(f_0)=\\
=1-\erf~\frac{\lambda(1)}{\sqrt{2}}+q\l[\erf~\frac{\lambda(1)}{\sqrt{2}} 
-\erf~\frac{\lambda(f_0)}{\sqrt{2}}\r]
\end{multline}
where $\lambda(f)$ is given by (\ref{lf}) with
$\ovl{f}=1-q(1-f_0)$. Eq. (\ref{cmax}) is given by
\begin{equation}\label{ccccc}
c=(1-q)c(1)+qf_0c(f_0)
\end{equation}
with $c(f)$ given by (\ref{cppp}), while, finally, the susceptibility
turns out to be expressed as
\begin{equation}\label{xxxxx}
\chi=(1-q)\chi(1)+q f_0 \chi(f_0)
\end{equation}
with $\chi(f)$ given by (\ref{chif}).

Equations (\ref{ccccc}) and (\ref{xxxxx}) must be solved
self-consistently and are valid as long as none of the assumptions
(\ref{tti}--\ref{wltm}) is violated. In the MG, ergodicity breaking is
related to the onset of anomalous response, i.e. to a divergence of
$\chi$. Working out (\ref{xxxxx}) explicitly, one finds the relation
\begin{equation}
\frac{\alpha\chi}{1+\chi}=(1-q)~\erf~\frac{\lambda(1)}{\sqrt{2}}
+q~\erf~\frac{\lambda(f_0)}{\sqrt{2}}
\end{equation}
$\chi$ is evidently finite for large $\alpha$. The critical values of
the control parameter where $\chi$ diverges are immediately obtained
as
\begin{equation}\label{accc}
\alpha_c(q,f_0)=(1-q)~\erf(x)+q~\erf(x\sqrt{f_0})
\end{equation}
where $x$ solves (\ref{ccccc}) at $\alpha_c$. After some algebra, the
latter condition is shown to be, in terms of $x$, equivalent to the
following:
\begin{multline}\label{accc2}
2\ovl{f}-(1-q)~\erf(x)-qf_0~\erf(x\sqrt{f_0})+\\
-\frac{1-q}{x\sqrt{\pi}}~e^{-x^2}-\frac{q}{x}\sqrt{\frac{f_0}{\pi}}~
e^{-f_0x^2}=0
\end{multline}
For $\alpha>\alpha_c$, the system is ergodic and admits a unique
stationary state. At $\alpha_c$, the integrated response diverges and
anomalous response sets in. For $\alpha<\alpha_c$, finally, ergodicity
is broken and the long-time limit of macroscopic quantities depends on
the initial conditions of the dynamics. Clearly, the limiting cases
$q=0$ and $f_0=1$ reproduce, as they should, the results of the
standard MG, while the limit $q=1$ brings us back to the standard case
after a trivial re-scaling. Notice that, as usual in MGs, $\chi$
diverges when $\alpha$ equals the fraction $1-\phi$ of fickle traders.

\begin{figure}[t]
\begin{center}
\includegraphics[width=8cm]{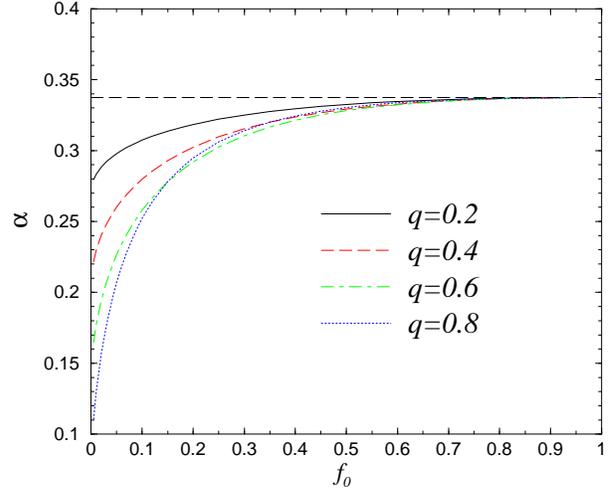}
\caption{\label{pd-1}Phase diagram of the multi-frequency MG with
bimodal frequency distribution. The curves $\alpha_c$ vs $f_0$
(trading frequency of occasional traders) are shown for different
values of the fraction $q$ of occasional traders. For
$\alpha>\alpha_c$ the system is ergodic. The horizontal dashed line at
$\alpha\simeq 0.3374$ marks the position of the critical point of the
standard MG and coincides with the critical line in the case $q=0$.}
\end{center}
\end{figure}

Numerical solution of (\ref{accc2}) and subsequent insertion of the
result into (\ref{accc}) leads to the $(f_0,\alpha)$ phase diagrams
shown in Fig.~\ref{pd-1} for different values of $q$.
For sufficiently small $f_0$, the critical point $\alpha_c$ decreases
as $q$ increases, signaling that the ergodic phase gets larger while
the efficient phase shrinks. As $f_0$ grows, $\alpha_c$ tends instead
to the standard MG value $0.3374\ldots$ for any $q$, as was
easily predictable. 

\begin{figure}[!]
\begin{center}
\includegraphics[width=8.5cm]{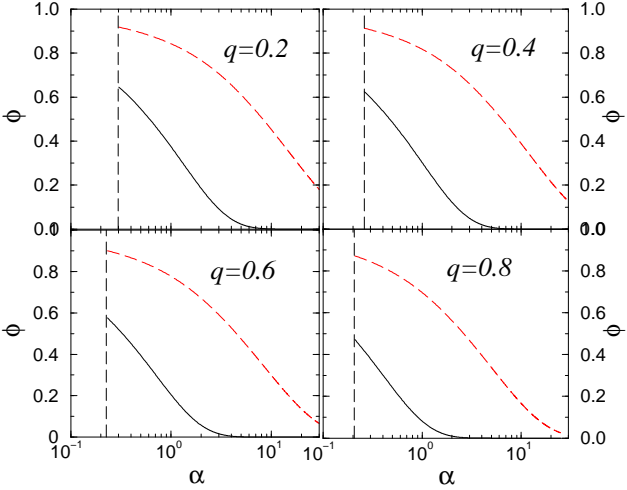}
\caption{\label{phi-1}Fraction $\phi(f)$ of agents with trading
frequency $f$ that are frozen for values of $q$ as reported in the
figures. The vertical dashed lines give the positions of the critical
points, below which (\ref{fir}) is violated and the theory is no
longer valid. The continuous (resp. dashed) line represents $\phi(1)$
(resp. $\phi(0.05)$).}
\end{center}
\end{figure}

Solving instead (\ref{phiphi}), (\ref{ccccc}), and (\ref{xxxxx}) one
obtains the behavior of the macroscopic order parameters and of the
fraction of frozen agents as a function of $\alpha$ upon varying $q$
and $f_0$. These quantities give an idea of the interplay between the
two types of agents. For sakes of conciseness and because we believe
it is the most interesting instance, we concentrate on the case where
the trading frequency of occasional traders is much lower than that of
regulars, that is we assume regulars and occasional traders operate on
widely different time scales. We take in particular
$f_0=0.05$. Results for $\phi(f)$ are shown in Fig.~\ref{phi-1}.
It is clear that occasional traders are more likely to freeze than
regular traders, as also argued in the static solution outlined in
\cite{mp}. This tendency is robust with respect to changes of $q$ and
$f_0$. It is interesting to notice that a larger fraction of
occasional traders freeze when regulars outnumber them.

In order to test our predictions, in particular for what concerns the
phase transition, and to understand further the combined effects of
regulars and irregulars on global efficiency in this simple case, we
can study the {\em volatility} matrix, defined as \cite{hc}
\begin{equation}\label{xi}
\mathsf{\Xi}=\frac{1}{2}~\mathsf{\Lambda}=\frac{1}{2}~
[(\mathsf{I}+\mathsf{G})^{-1}(\mathsf{F}+\mathsf{C})
(\mathsf{I}+\mathsf{G}^T)^{-1}]
\end{equation}
$\mathsf{\Xi}$ has two important properties. First, the magnitude of
global fluctuations, i.e. $\sigma^2=\timeavg{A^2}$, which serves as a
measure of global efficiency (the smaller is $\sigma^2$, the more
efficient is the allocation of resources, that is, the smaller the
amount of resource that is wasted), is given by
\begin{equation}\label{ippo}
\sigma^2=\lim_{t\to\infty}\Xi(t,t)
\end{equation}
Second, the quantity $H=(1/P)\sum_{\mu=1,P}\timeavg{A|\mu}^2$, where
\begin{equation}
\timeavg{A|\nu}=\lim_{L\to\infty}\frac{1}{L-L_{\rm eq}}\sum_{n=L_{{\rm
eq}},L} A(n)\delta_{\mu(n),\nu}
\end{equation}
that quantifies the informational efficiency of the system (when $H=0$
the minority action is not predictable on the basis of the state of
the world $\mu(n)$ alone and the system is efficient, in the sense
that it does not create information an external agent could exploit to
have a gain) is roughly given by the persistent part of
$\mathsf{\Xi}$. Both these properties can be proved under general
conditions, as done for instance in \cite{hc}.

\begin{figure}[!]
\begin{center}
\includegraphics[width=8cm]{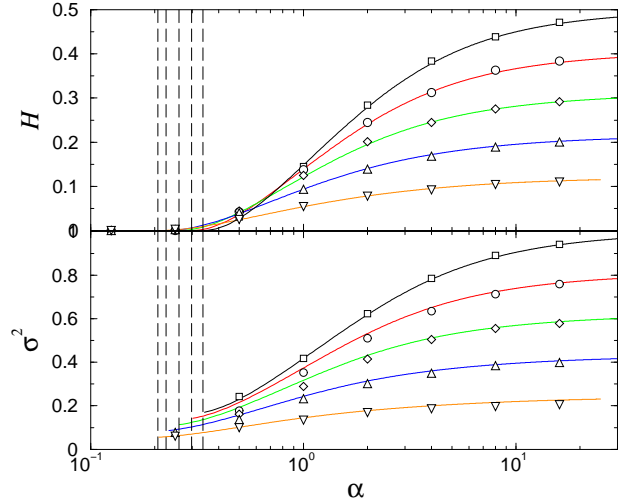}
\caption{\label{Hs2-1}Behavior of $H$ and $\sigma^2$ vs $\alpha$ at
fixed $f_0=0.05$ for $q=0,0.2,0.4,0.6,0.8$ (continuous curves, top to
bottom). The vertical dashed lines mark the positions of the critical
points. In the non-ergodic regime $\alpha<\alpha_c$ the theory is no
longer valid. Markers denote the results of computer simulations of a
system of $256$ agents for the same values of $f_0$ and $q$, averaged
over $100$ disorder samples.}
\end{center}
\end{figure}

$H$ is easily seen from (\ref{xi}) to be given by
\begin{equation}\label{acca}
H=\frac{\ovl{f}+c}{2(1+\chi)^2}
\end{equation}
and tends to $\ovl{f}/2$ for $\alpha\to\infty$. $\sigma^2$ can instead
be approximated with
\begin{equation}\label{sgama}
\sigma^2=\frac{\ovl{f}+\int f q(f)\phi(f)df}{2(1+\chi)^2}
+\frac{1}{2}\l[\ovl{f}-\int f q(f)\phi(f)df\r]
\end{equation}
and tends to $\ovl{f}$ for $\alpha\to\infty$. The proof of
(\ref{sgama}), which requires an approximate evaluation of
(\ref{ippo}), is a slight modification of the standard case, reported
in \cite{hc}, and we will not report it here. For the present model,
we have obtained $\sigma^2$ and $H$ from numerical simulations
(performed with flat or unbiased initial conditions $y_i(0)=0$ for all
$i$). The comparison with analytical predictions can be seen in
Fig.~\ref{Hs2-1}.
As in the standard MG, the ergodicity breaking transition coincides
with a transition between an informationally efficient phase with
$H=0$ ($\alpha<\alpha_c$) and an informationally inefficient one with
$H>0$ ($\alpha>\alpha_c$). $H$ vanishes at at $\alpha_c$, where
$\chi\to\infty$, and decreases as $q$ increases. At the same time,
$\sigma^2$ decreases in vicinity of the critical point, signaling an
increased degree of cooperation.

\section{Power-law frequency distribution}

\begin{figure}[!]
\begin{center}
\includegraphics[width=8cm]{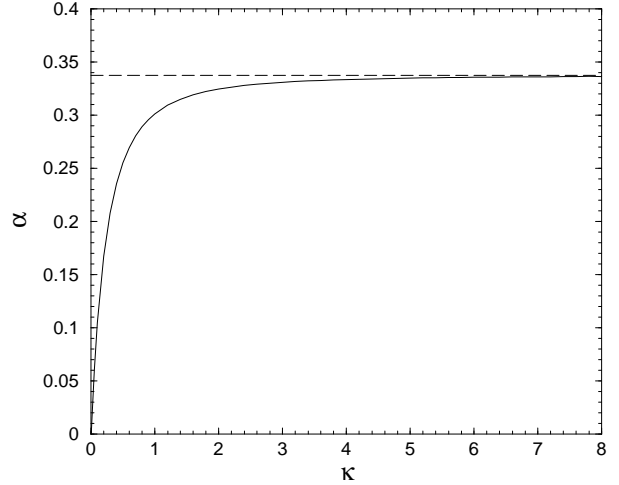}
\caption{\label{pd-2}Phase diagram of the multi-frequency MG with
power-law frequency distribution. The curve $\alpha_c$ vs $\kappa$ is
shown. For $\alpha>\alpha_c$ the system is ergodic and informationally
inefficient. The horizontal dashed line at $\alpha\simeq 0.3374$ marks
the position of the critical point of the standard MG.}
\end{center}
\end{figure}

Let us now turn our attention to the case where
\begin{equation}
q(f)=\kappa f^{\kappa-1},~~~~~~\ovl{f}=\frac{\kappa}{\kappa+1}
\end{equation}
where the frequency distribution is scale-free (i.e., there is no
single characteristic time scale). The standard instance is recovered
in the limit $\kappa\to\infty$, in which $q(f)\to\delta(f-1)$.  This
problem was treated statically in \cite{mp}, so we will limit
ourselves to calculating the phase diagram and discussing its
qualitative properties. The general dynamical solution is now
slightly more complicated than the bimodal case, however it is still
possible to write down closed equations for the relevant
parameters. The behavior of macroscopic observables turns however out
to be similar to the bimodal case. Let us first notice that, since
\begin{equation}
1-\phi=\kappa\int_0^1 f^{\kappa-1}\erf~\frac{\lambda(f)}{\sqrt{2}}df
\end{equation}
and since, from (\ref{chif}),
\begin{equation}
\frac{\alpha\chi}{1+\chi}=1-\phi
\end{equation}
we immediately obtain the expression for the critical line signaling
ergodicity breaking in the $(\kappa,\alpha)$ space, namely
\begin{equation}\label{alfacc}
\alpha_c(\kappa)=\kappa\int_0^1 f^{\kappa-1}\erf(x\sqrt{f})df
\end{equation}
where again $x$ solves the equation for $c$, namely
\begin{equation}\label{cici}
c=\kappa\int f^\kappa c(f)df 
\end{equation}
at $\alpha_c$. One can see by a short calculation that in terms of $x$
and for $\alpha=\alpha_c$, Eq. (\ref{alfacc}), equation (\ref{cici})
reads
\begin{equation}\label{equa}
\frac{2\kappa}{1+\kappa}-\kappa\int_0^1 f^{\kappa-1}\l[f~\erf(x\sqrt{f})
+\frac{\sqrt{f}}{x\sqrt{\pi}}~e^{-fx^2}\r]df=0
\end{equation}
When $\kappa\to\infty$, the standard picture is recovered. Using a
more compact notation, (\ref{equa}) be re-cast as
\begin{equation}\label{eqa}
\frac{\kappa}{1+\kappa}[2-\erf(x)]+\frac{\mathcal{F}(\kappa,x;3/2)}{1+\kappa}
-\mathcal{F}(\kappa,x;1/2)=0
\end{equation}
where we introduced the shorthand
\begin{equation}
\mathcal{F}(\kappa,x;a)=\frac{\kappa}{x^{2(\kappa+1)}\sqrt{\pi}}
\l[\Gamma(\kappa+a)-\Gamma^u(\kappa+a,x^2)\r]
\end{equation}
with $\Gamma(\cdot)$ and $\Gamma^u(\cdot,\cdot)$ the Euler Gamma
function and the upper incomplete Gamma function,
respectively. (\ref{eqa}) can be solved numerically for $x$ upon
varying $\kappa$. Subsequently, the corresponding value of $\alpha_c$
can be calculated from (\ref{alfacc}):
\begin{equation}
\alpha_c(\kappa)=\erf(x)-\frac{x^2}{\kappa}\mathcal{F}(\kappa,x;1/2)
\end{equation}
The resulting phase diagram is shown in Fig.~\ref{pd-2}.
For large $\kappa$, the standard MG picture is recovered, as
expected. As $\kappa$ decreases, $\alpha_c$ also decreases, proving
that the fraction of active agents is also getting smaller and
smaller. $\alpha_c$ finally tends to zero when $\kappa\to 0$. While
this picture is qualitatively identical to that derived in \cite{mp},
there are certain numerical differences as in our solution
$\alpha_c(\kappa)$ seems to approach the standard limit
$\alpha_c(\infty)\simeq 0.3374$ slightly faster than in the replica
solution. However a direct comparison could be misleading since the
order parameters are defined differently in the two cases. This point
might deserve further investigation. The fraction $\phi$ of frozen
agents behaves similarly to what shown in Fig.~\ref{phi-1}, in that
agents trading less frequently are more likely to be frozen, as also
discussed in \cite{mp}. Clearly, the formulas derived in the previous
section for $\sigma^2$ and $H$ can be applied, {\em mutatis mutandis},
also in this case.

\section{CMG with finite learning rates}

We now add one layer of difficulty by introducing a learning rate in
the individual agents' dynamics of strategy selection, as first done
for the standard model in \cite{cggs}. Up to now, agents have strictly
followed their best performing strategy at each round, as encoded in
the deterministic rule
\begin{equation}\label{strict}
s_i(n)=\sign[y_i(n)]
\end{equation}
Now we account for the possibility that at each time step the value of
$s_i(n)$ is established by a probabilistic rule. The simplest choice
for the latter is\footnote{We use for the learning rate the customary
symbol $\Gamma$. The reader is warned that it is different from the
$\Gamma$ function appearing in Sec.~4.}
\begin{equation}
\prob\{s_i(n)=\pm 1\}\sim e^{\pm \Gamma_i y_i(n)}
\end{equation}
which turns (\ref{strict}) into
\begin{equation}\label{adn}
s_i(n|\zeta_i(n),\Gamma_i)=\sign[y_i(n)+\zeta_i(n)/\Gamma_i]
\end{equation}
where the $\zeta_i$'s are i.i.d. Gaussian r.v.'s with probability
density $p(\zeta)=[1-\tanh^2(\zeta)]/2$ (zero average, unit
variance). The positive constants $\Gamma_i$ can be interpreted as
`learning rates', upon varying which one interpolates between the
deterministic rule (\ref{strict}), that is recovered for
$\Gamma_i\to\infty$, and a fully randomized rule, corresponding to
$\Gamma_i=0$. The dynamics of the standard model with finite learning
rates was tackled in \cite{chs}, where the phrase `additive decision
noise' was introduced to describe the situation of (\ref{adn}). Notice
that this type of noise affects fickle traders only, because for
`frozen' traders $|y_i(n)|\to\infty$, hence it is to be expected that
the critical points obtained for different trading frequencies do not
change. A more complicated situation is that of `multiplicative
decision noise', first introduced in \cite{gms}, where instead of
(\ref{adn}) one has
\begin{equation}\label{mdn}
s_i(n|\zeta_i(n),\Gamma_i)=\sign[y_i(n)(1+\zeta_i(n)/\Gamma_i)]
\end{equation}
which corresponds to
\begin{equation}
\prob\{s_i(n)=\pm 1\}\sim e^{\pm \Gamma_i\sign[y_i(n)]}
\end{equation}
This choice clearly affects also frozen agents. 

The dynamical theory of this case is a straightforward modification of
that constructed for the deterministic case. In general, with
$s(n|\zeta_i(n),\Gamma_i)$, the dynamics (\ref{batch}) becomes
\begin{equation}
y_i(t+1)=y_i(t)-h_i-\sum_{j=1}^N J_{ij}s_j(t|\zeta_j(t),\Gamma_j)
\end{equation}
Again, one can write down its generating functional, taking into
account the fact that this time the calculation of one-step transition
probabilities, the product of which is used to compute averages over
paths, requires carrying out an average over the $\zeta_i$'s (we
denote this operation by $\zavg{\cdots}$):
\begin{multline}
Z[\boldsymbol{\psi}]=\pathavg{e^{\ii\sum_{it}y_i(t)\psi_i(t)}}=\\
=\int p(\boldsymbol{y}(0))~ e^{\ii\sum_{it}\widehat{y}_i(t)
[y_i(t+1)-y_i(t)+h_i-\theta_i(t)]+y_i(t)\psi_i(t)}\times\\\times
\zavg{e^{\ii\sum_{it}\widehat{y}_i(t)\sum_j
J_{ij}s_j(t|\zeta_j(t),\Gamma_j)}}
\prod_{it}[dy_i(t)d\widehat{y}_i(t)/(2\pi)]
\end{multline}
Again, averaging over the quenched disorder leads to a non-Markovian
process that now describes a single effective agent trading with
frequency $f$ {\em and} learning at rate $\Gamma$. And again, we skip
all formalities of the derivation of this process and focus on its
long-time properties. The analog of (\ref{esap}) in this case is
\begin{equation}\label{esap2}
y(t+1)=y(t)+\sum_{t'\leq t}R(t,t')s(t'|\zeta(t'),\Gamma)
+\theta(t)+\sqrt{\alpha f}z(t)
\end{equation}
(\ref{corr}) and (\ref{rsi}) are still formally valid, but the
correlation and response functions $C(t,t')$ and $G(t,t')$ are defined
respectively as
\begin{gather}
C(t,t')=\avg{s(t|\zeta(t),\Gamma)s(t'|\zeta(t'),\Gamma)}_*\\
G(t,t')=\frac{\partial}{\partial\theta(t')}\avg{s(t|\zeta(t),\Gamma)}_*
\end{gather}
where the average $\avg{\cdots}_*$, formerly given by (\ref{star}),
involves now also averages over $\boldsymbol{z}$ and $\Gamma$. In
particular, if we denote by $w(\Gamma)$ the probability density of
$\Gamma$, we have
\begin{multline}\label{starr}
\avg{\cdots}_*=\int_0^1 df f q(f)\int_0^\infty d\Gamma
w(\Gamma)\int\prod_t\frac{dy(t)d\widehat{y}(t)}{2\pi}~\times\\ \times
\Big[(\cdots)~e^{-\frac{\alpha
f}{2}\sum_{tt'}\widehat{y}(t)\Lambda(t,t')\widehat{y}(t')}\times\\
\times e^{\ii\sum_t
\widehat{y}(t)[y(t+1)-y(t)-\theta(t)-\sum_{t'}R(t,t')s(t'|\zeta(t'),\Gamma)]}
\Big]_{\boldsymbol{\zeta}}
\end{multline}
As before, $C(t,t')$ and $G(t,t')$ represent the disorder-averaged
correlation and response functions of the original system. 

Proceeding as done in Sec.~2 by assuming the validity of
(\ref{tti}--\ref{wltm}), one can derive an equation for the quantity
$\widetilde{y}=\lim_{t\to\infty}y(t)/t$, which reads
\begin{equation}\label{longo2}
\frac{\widetilde{y}}{\sqrt{\alpha f}}=-\gamma m(\Gamma)+z
\end{equation}
where all quantities have the same meaning as in (\ref{longo}) except
that now $m(\Gamma)$ is the time-average of
$s(t|\zeta(t),\Gamma)$. This change presents no additional
difficulty. In the next section we will analyze the stationary states
and the phase diagrams resulting from (\ref{adn}) and (\ref{mdn})
separately.

\section{Additive and multiplicative decision noises}

In the case of (\ref{adn}), one immediately sees that the conditions
for an agent to be frozen or fickle are identical to those given in
Sec.~2, hence the picture emerging from different trading frequencies
(in particular the expressions for $\chi$, $c$ and $\phi$) is left
unchanged since $\Gamma$ can be integrated out. In particular, the
phase diagrams presented in Sec.~3 and~4 for bimodal and power-law
frequency distribution remain valid. Hence, as in the standard batch
MG, the introduction of additive noise does not affect the macroscopic
properties of the multi-frequency model in the ergodic regime. (On the
other hand, the introduction of $\Gamma$ does affect the stationary
states of the standard model in the non-ergodic regime, as first shown
by computer simulations in \cite{cggs} and by analytical arguments in
\cite{mc}.)

Finally, we turn to the case of multiplicative decision noise
(\ref{mdn}). Following \cite{chs}, it is convenient to switch to the
inverse learning rate $T=1/\Gamma$ and to introduce the quantity
\begin{equation}\label{acc}
h(T)=\int\sign(1+T\zeta)p(\zeta)d\zeta
\end{equation}
which has the obvious properties $h(T)\in[0,1]$ with $h(0)=1$ and
$h(\infty)=0$.  Using this, one concludes from (\ref{longo2}) with
$m(\Gamma)\equiv m(T)=\sign[\widetilde{y}]h(T)$ that frozen agents
with $\sign[\widetilde{y}]=\pm 1$ have $m(T)=\pm h(T)$, respectively,
and occur for $|z|>\gamma h(T)$, whereas fickle agents have
$m(T)=z/\gamma$ and occur for $|z|<\gamma h(T)$.

In order to make contact with the quantities calculated in the
deterministic decision-making case, we compute the autocorrelation of
the effective trader with trading frequency $f$, $c(f)$, which can as
usual be obtained by separating the contributions of frozen and fickle
agents, this time performing an additional average over $T$. For
simplicity, one can pass from the variable $T$ with probability
density $\widetilde{w}(T)$ to $h$, Eq. (\ref{acc}), with probability
density
\begin{equation}
\rho(h)=\int_0^\infty \delta\l[h-\int\sign(1+T\zeta)p(\zeta)d\zeta\r]~
\widetilde{w}(T)dT
\end{equation}
so that
\begin{equation}
c(f)=\int[\avg{\theta(|z|-\gamma h) h^2}+\avg{\theta(\gamma
h-|z|)(z/\gamma)^2}]~\rho(h)dh
\end{equation}
where $\avg{\cdots}$ denotes an average over the Gaussian r.v. $z$
with zero average and variance $\avg{z^2}=(\ovl{f}+c)/(1+\chi)^2$,
where
\begin{equation}\label{impa}
c=\int f q(f) c(f) df
\end{equation}
is the global persistent autocorrelation. The fraction of frozen
agents reads
\begin{equation}
\phi=\int df q(f)\int dh \rho(h)\avg{\theta(|z|-\gamma h)}
\end{equation}
where it should be kept in mind that $\gamma$ depends on $f$
($\gamma=\sqrt{\alpha f}/(1+\chi)$).

An expression for the total susceptibility $\chi$ can be found
starting from (\ref{chippp}) and averaging properly over $f$ and $h$:
\begin{equation}
\frac{\alpha\chi}{1+\chi}=1-\int df q(f)\int dh
\rho(h)\avg{\theta(|z|-\gamma h)}\equiv 1-\phi
\end{equation}
so that as usual $\chi$ diverges when the fraction of fickle agents
equals $\alpha$. We have thus come to the following expression for the
critical line:
\begin{equation}
\alpha_c[q(f),w(\Gamma)]=\int df q(f)\int dh \rho(h)~\erf~(xh\sqrt{f})
\end{equation}
where we have emphasized that $\alpha_c$ now depends on the trading
frequency distribution $q(f)$ and on the distribution of learning
rates $w(\Gamma)$ (via $\widetilde{w}(T)$ and $\rho(h)$). Again, the
value of $x$ must be obtained by solving (\ref{impa}) at $\alpha_c$.

\begin{figure}[!]
\begin{center}
\includegraphics[width=8.5cm]{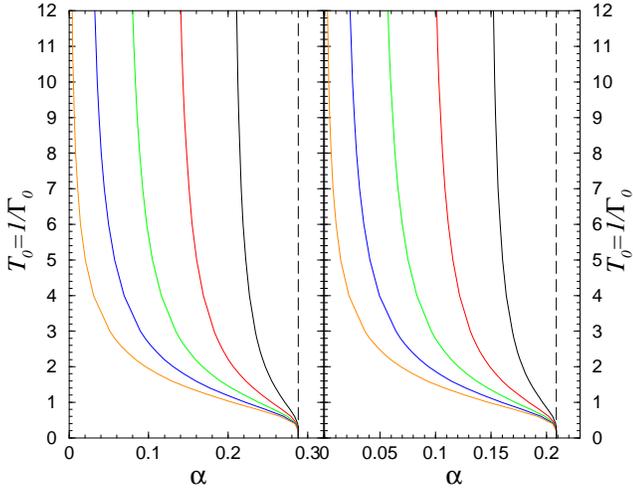}
\caption{\label{pd-Gamma}Phase diagram of the multi-frequency MG with
stochastic decision-making, with bimodal distributions for frequencies
and learning rates, with parameter $f_0=0.05$. Results are shown for
$q=0.25$ (left panel) and $q=0.75$ (right panel). In each panel,
continuous curves denote the critical lines for multiplicative
decision noise for $\epsilon=0.2,0.4,0.6,0.8,1$ (right to left). The
dashed vertical line marks the position of the critical point for the
CMG with deterministic decision-making, which coincides with the
critical line for the corresponding CMG with additive decision noise.}
\end{center}
\end{figure}

For the sake of brevity, we specialize the above theory only to the
simplest cases where both $f$ and $T$ have binomial distributions:
\begin{gather}
q(f)=(1-q)\delta(f-1)+q\delta(f-f_0)\\
\widetilde{w}(T)=(1-\epsilon)\delta(T)+\epsilon\delta(T-T_0)
\end{gather}
with $q$, $f_0$ and $\epsilon$ real numbers belonging to the interval
$[0,1]$. This describes a mixed population of frequent and occasional
(trading frequency $f_0$) traders, with either deterministic decision
making or stochastic decision making with learning rate
$\Gamma_0=1/T_0$. Notice that trading frequency and learning rate are
treated as independent variables (this seems to the author a rather
strong and possibly unrealistic assumption). Under these hypotheses,
one has
\begin{equation}
\rho(h)=(1-\epsilon)\delta(h-1)+\epsilon\delta(h-h_0),~~~~ h_0=h(T_0)
\end{equation}
Introducing again $\lambda(f)=\sqrt{\frac{\alpha f}{\ovl{f}+c}}$ and
setting
\begin{multline}
c(f,h)=h^2\l[1-\erf~\frac{\lambda(f)h}{\sqrt{2}}\r]+\\
+\frac{1}{\lambda(f)^2}~\erf~\frac{\lambda(f)h}{\sqrt{2}}-
\frac{h}{\lambda(f)}\sqrt{\frac{2}{\pi}}~e^{-\lambda(f)^2 h^2/2}
\end{multline}
one finds, for $c$,
\begin{multline}\label{cl}
c=(1-q)(1-\epsilon)~c(1,1)+(1-q)\epsilon ~c(1,h_0)+\\
+f_0q(1-\epsilon)~c(f_0,1)+f_0 q\epsilon ~c(f_0,h_0)
\end{multline}
and for $\alpha_c$, which now depends on the population fractions $q$
and $\epsilon$ and on $f_0$ and $h_0$,
\begin{multline}
\alpha_c=(1-q)(1-\epsilon)~\erf(x)+(1-q)\epsilon~\erf(xh_0)+\\
+q(1-\epsilon)~\erf(x\sqrt{f_0})+q\epsilon~\erf(x h_0\sqrt{f_0})
\end{multline}
When $\alpha=\alpha_c$, (\ref{cl}) takes a simple but rather lengthy
expression which we do not report. Ultimately, all equations can be
solved numerically. As already done in Sec.~3, we show the phase
diagram for the case where regular and occasional traders operate on
very different time scales and again choose $f_0=0.05$ (see
Fig.~\ref{pd-Gamma}).
One can clearly see that the phase structure of the MG with
multiplicative noise \cite{chs} is preserved, the main effect being a
shift of the position of the critical lines in the $(\alpha,\Gamma_0$)
space to lower values of $\alpha$ (signaling again a reduction of the
informationally efficient phase).

\section{Summary and discussion}

In summary, the MG with agents trading at different frequencies and
learning at different rates has been analyzed using
generating-functional techniques. The dynamical critical behavior of
the model is not altered by the introduction of multiple time scales
for trading, but the specific phase diagram turns out to depend on the
details of the underlying frequency and learning rate distribution. We
have derived the phase structure for a system of agents with
deterministic decision-making when the frequency distribution is a
simple bimodal and when it has a power-law form. For the case of
stochastic decision-making, i.e. finite learning rates, the phase
diagram has been computed explicitly for a double bimodal distribution
of frequencies and learning rates, describing fast and slow traders
with a heterogeneous distribution of stochasticity in their learning
process. In accordance with \cite{mp}, we found that occasional
traders display a stronger tendency to freeze, namely to use just one
of their strategies, than regular traders, thus contributing and
additional signal to unfrozen players. To the latter, thus mostly to
regular traders, is ultimately ascribable the building up of
fluctuations (via freezing) and thus the decrease of global
efficiency.

On a purely technical level, it is clear that combining static and
dynamical methods of statistical mechanics of disordered systems leads
to a very satisfactory understanding of the the macroscopic properties
of MG-based market models. On the other hand, MGs are simple enough to
hope they can serve as the elementary building blocks of more
complicated and realistic agent-based market models, with different
types of agents that are heterogeneous in strategies, beliefs,
learning rates, market-impact evaluation, etc., and where the plethora
of macroscopic phenomena observed in real systems could be analyzed
and traced back to the individual laws of motion. Present works
concerning the macroscopic properties of different versions of the MG
should be seen as the necessary preliminary work to develop a full
understanding at later stages.

Indeed, several directions are open for further research. For example,
it would be interesting to relate $\Gamma$ and $f$ by letting agents
with a smaller learning rate trade more (or less) frequently. This
could also be done by assuming that $\Gamma$ and $f$ are not
independent qualities, as done in Sec.~6 of the present work, but come
with a joint probability density. Another worthy extension could
involve the introduction of (using the terminology of \cite{cmz})
`producers', i.e. agents who have just one strategy at their
disposal. It is likely that they could again affect the phase
structure.

\bigskip

\textbf{Acknowledgments.} I am grateful to A. Cavagna, A.C.C. Coolen,
I. Giardina, J.A.F. Heimel, M. Marsili and G. Mosetti for
encouragement and many useful discussions and precious suggestions.

\end{document}